# Anisotropic Young-Laplace-equation provides insight into tissue growth.

Peter Fratzl[1], F. Dieter Fischer[2], Gerald A. Zickler[2], John W.C. Dunlop[3]

[1]Max Planck Institute of Colloids and Interfaces, Department of Biomaterials, Potsdam Science Park, 14476 Potsdam-Golm, Germany

[2]Montanuniversität Leoben, Institute of Mechanics, 8700 Leoben, Austria

[3]Morphophysics Group, Department of the Chemistry and Physics of Materials, University of Salzburg, 5020 Salzburg, Austria

*corresponding authors: Peter Fratzl and John W.C. Dunlop

**Email:** fratzl@mpikg.mpg.de, john.dunlop@plus.ac.at

**This PDF file includes:**

>Main Text
>Figures 1 to 4

**Abstract**

Growing tissues are highly dynamic, flowing on sufficiently long time-scales due to cell proliferation, migration and tissue remodeling. As a consequence, living tissues can be approximated as liquids. This means the shape of microtissues is governed by a surface stress state as in fluid droplets. Recent work showed that cells in the near-surface region of fibroblastic or osteoblastic microtissues contract with highly oriented actin filaments, thus making the surface stress state anisotropic, in contrast to what is expected for an isotropic fluid. Here, we extend the Young-Laplace law to include mechanical anisotropy of the surface. We then take this into account to determine equilibrium shapes of rotationally symmetric bodies subjected to anisotropic surface stress states and derive a family of surfaces of revolution in analogy to the Delaunay surfaces. A comparison with recently published shapes of microtissues shows that this theory accurately predicts both the surface shape and the direction of the actin filaments in the surface. The anisotropic version of the Young-Laplace law might help describing the growth of microtissues and also predicts the shape of fluid bodies with highly anisotropic surface properties.

**Main Text**

**Introduction**

Surface stresses have long been known to determine the shape of fluid bodies. The general idea is that the minimization of the total surface energy leads to droplet shapes with constant mean curvature, whereby the associated surface stresses counterbalance a pressure inside the fluid droplet in a way described by the Young-Laplace equation [1, 2]. More than hundred years ago, D'Arcy Thompson proposed that surface tension could also play a role in the morphogenesis of biological organisms [3]. Indeed, growing tissues where cells permanently divide and rearrange behave – over sufficiently long time-scales – as fluids, where shear stresses relax over time and only isostatic pressure remains. D'Arcy Thompson's ideas have meanwhile found their way into modern developmental biology, where the importance of mechanical forces for morphogenesis receive increasing attention [4, 5]. While a fluid droplet subjected only to surface energy would converge to the simplest shape with constant mean curvature, a sphere, the combination with the adherence to another wettable surface may lead to the emergence of arbitrarily complex shapes depending on the boundary conditions [6]. It has also been shown that surface energy not only influences the shape of growing tissues but also their growth kinetics [7], a phenomenon that can be rationalized through simple models of tissue growth [8]. Such concepts have significant practical implications for the optimal design of scaffolds for tissue engineering [9, 10], for example.

Surface energy concepts have been successfully applied especially to the development of epithelial layers which are naturally two-dimensional tissues [4, 5]. But it has also been shown that in-vitro grown microtissues based on connective-tissue-forming fibroblasts develop a contractile layer at the surface of the growing tissue where cells temporarily turn into myofibroblasts [11] capable of generating surface stresses. The shape of the resulting microtissues are fully compatible with constant curvature surfaces (Fig 1A). Recently it was shown that microtissues with cylindrical symmetry grown in-vitro evolve into shapes with cylindrical symmetry and constant mean curvature [12], known as Delaunay surfaces [13, 14] (Fig 1 B). This is to be expected from the Young-Laplace equation that assumes isotropic, in-plane surface mechanical properties. However, the same recent work also revealed a puzzling observation that remains unexplained: fluorescently-stained, contractile actin filaments in the microtissue surface turn out to be highly aligned and follow helical paths around the tissue [12] (Fig. 1C) and indicate, therefore, that the surface does not have isotropic in-plane properties.

In this paper, we aim to explore the surprising finding that tissues may grow into shapes close to constant curvature surfaces, despite having extremely anisotropic in-plane mechanical surface properties, in contrast to the assumption that the traditional Young-Laplace law holds. For this, we first derive a simple extension of the Young-Laplace law, assuming that the surface properties are maximally anisotropic, with vanishing in-plane elastic modulus in the direction perpendicular to the fibers in the surface. Note that this type of anisotropy is not the same as what is studied for the surface of crystals [15] or liquid crystals [16]. In those cases, the underlying bulk material has itself anisotropic properties. We also assume that the fibrous reinforcement remains bound to the surface in contrast to growing fibers constrained within a volume [17]. The case that we consider is inherently simpler and consists of an isotropic fluid bounded by a surface layer, modeled as a shell that consists of locally parallel fibers. Considering that the isostatic pressure inside the fluid is in equilibrium with the force in the surface fibers, the Young Laplace equation is shown to be replaced by two simple conditions. We then consider the special case of volumes with rotational symmetry around an axis and derive the corresponding equilibrium shapes that are distinct from Delaunay surfaces corresponding to isotropic surface properties. The resulting shapes all show spiral arrangements of fibers in the surface and for some special situations, the tissue shapes are qualitatively close to Delaunay shapes. This is finally discussed in terms of the known experimental observations [12].

**Modification of the Young-Laplace equation for a fiber surface**

*Surface stress and pressure*

The Young-Laplace equation was derived by both Young and Laplace in 1805 [1, 2]. It describes how the pressure difference acting on the surface of a thin shell (e.g. a drop), $p$ is related to the surface stress $\gamma_s$ (or surface tension, dimension [N/m]) via the prominent relation

$$p = -\gamma_s(1/R_1 + 1/R_2) = -2\gamma_s H. \qquad [1]$$

The surface stress $\gamma_s$ may also consist of several contributions, see e.g. [16, 18]. Since $\gamma_s$ is assumed to be independent of any direction, often the term "isotropic surface stress" is used.

Both quantities, $R_1$ and $R_2$ in Eq. **1**, are the signed principal radii of curvature; the expression $(1/R_1 + 1/R_2)$ is thus twice the mean curvature $H$. This equation is a special formulation of the equilibrium equation for a membrane (see e.g. [19-23]) via the signed principal curvatures $1/R_1$ and $1/R_2$ with corresponding membrane forces $N_1, N_2$, reading as

$$p = \left(\frac{N_1}{R_1} + \frac{N_2}{R_2}\right). \qquad [2]$$

$N_1$ and $N_2$ are the tensile membrane forces measured in the direction of the orthogonal principal curvature lines, denoted by the unit vectors $\boldsymbol{e_1}$ and $\boldsymbol{e_2}$. The quantity $p$ is the pressure difference across the membrane measured in the direction of $\boldsymbol{e_1} \times \boldsymbol{e_2} = \boldsymbol{e_3}$. This equilibrium equation is general and does not rely on the symmetry of any particular configurations, such as axisymmetric and spherically symmetric shells for example. It should be mentioned that $N_1$ and $N_2$, instead of the constant surface stress $\gamma_s$ were sometimes denoted in the literature as "anisotropic surface tensions" (e.g. [24]). From the point-of-view of the mechanics of materials, the term "anisotropic" is, however, assigned to the material behavior.

The Young-Laplace equation Eq. **1** has motivated the mathematical community to look for surfaces with constant mean curvature $H$, and thus constant pressure difference $p$. In 1841, Delaunay showed that the following surfaces of revolution namely; spheres, cylinders, nodoids, catenoids and unduloids, all have constant mean curvature $H$ and thus satisfy the Young-Laplace equation [13]. The Young-Laplace equation has previously also been used to describe tissue growth [25].

*In-plane anisotropic properties*

Here we consider an isotropic fluid bounded by a surface layer modeled as a thin, unloaded shell reinforced with long fibers. A pressure difference $p$ is applied across this surface layer. The fibers then can be considered as the load carrying components of an anisotropic composite, for example the actin stress fibers in the tissues shown of Fig 1C [12]. Static equilibrium enforces that the direction $e_T$ of any fiber is colinear with the (local) resultant membrane force (due to $N_1$ and $N_2$) in the tangent plane to a material point of the shell. We further define the local signed curvature $1/R_T$ of the fiber in direction of $e_T$. It becomes immediately clear that for equilibrium the fiber needs to fulfill two conditions (Fig 1F). The first one is due to the fact that the force resulting from a tension on the fiber will be within the osculating plane of the fiber at this point. This force (denoted $s$ in Fig. 1F) will be along the principle normal to the curve, $e_p$, a unit vector perpendicular to the curve and pointing towards the local center of curvature and, therefore, lying within the osculating plane of the curve. The pressure difference across the surface results in a force that acts along the normal direction to the surface $e_n$. In general, this is not be parallel to the principal normal to the curve representing the fiber at the same point (Fig. 1F). Therefore, mechanical equilibrium requires that $e_n = e_p$. The second condition links the magnitude of the fiber load with the pressure difference across the surface. A representative membrane force $\sigma_T \cdot d$ is assigned to the shell/fiber system with an average thickness $d$ and an average load stress $\sigma_T$. According to Eq. **2** the local equilibrium between the fiber-shell system and the pressure difference $p$ enforces

$$p = \frac{d}{R_T} \sigma_T. \qquad \text{[3]}$$

Indeed, the membrane force perpendicular to the fiber is zero and along the fiber it is $d\,\sigma_T$, so that Eq. **2** reduces directly to Eq. **3**. Given that the pressure inside the volume is constant (as it should be for an isotropic fluid) and that the load along a fiber should also be constant, the requirement for a constant mean curvature that results from the Young Laplace equation (Eq. **1**) needs to be replaced by two conditions:

$$e_n = e_p \qquad \text{[4a]}$$
$$R_T = \text{constant}. \qquad \text{[4b]}$$

The first condition ensures that the resultant force on the volume generated by the fiber tension is parallel to the surface normal (and, thus, able to compensate the internal pressure). The second condition ensures that the force along the fiber is constant.

**Fiber-Supported Surfaces of Revolution – Generalization of Delaunay Surfaces**

In order to get a geometric understanding of the requirements of Eqs. **4a,b**, we analyze surfaces of revolution that fulfil these conditions. Since the experiment data reported in Fig. 1 were also obtained with tissues growing on surfaces of revolution, this will allow us a direct comparison with these experiments. In principle, however, the generalization of constant mean curvature surfaces as defined by Eqs. **4a,b** does not need to be rotationally symmetric, depending on boundary conditions.

*Calculating the equilibrium for fibers on surfaces of revolution*

We consider a surface of revolution $X$ given in a Cartesian coordinate system by the coordinates $x, y, z$ as products of the shape function $g(z)$ and the polar angle $\theta$ in the $x - y$ plane. Moreover, we assume that this surface consists of fibers that are positioned around the $z$-axis according to a function $\theta = \theta(z) + \theta_0$. The angle $\theta_0$ at $z = 0$ indicates the starting point of any particular fiber. Making use of the rotational symmetry, we restrict our analyses to the fiber where $\theta(0) = 0$. Therefore, we can describe the surface of the system in vector form $X(\theta, z)$ depending on only two-coordinates, $\theta$ and $z$.

$$X(\theta, z) = \begin{pmatrix} x \\ y \\ z \end{pmatrix} = \begin{pmatrix} g(z)\cos\theta \\ g(z)\sin\theta \\ z \end{pmatrix}, \quad [5]$$

Using standard differential geometry (see e.g. [26, 27]) we can determine the unit normal vector of the fiber $e_p$ (that is the unit normal vector lying within the osculating plane and along which the resulting force onto the surface will be directed).

$$e_p = \frac{1}{\sqrt{x''^2+y''^2+(y''x'-x''y')^2}\sqrt{x'^2+y'^2+1}} \cdot \begin{pmatrix} x''(1+y'^2) - y''x'y' \\ y''(1+x'^2) - x''x'y' \\ -(x'x'' + y'y'') \end{pmatrix}. \quad [6]$$

Note that primes refer to derivatives with respect to $z$. Likewise, the unit surface normal, $e_n$, is given by

$$e_n = \frac{1}{\sqrt{x^2(1+x'^2)+y^2(1+y'^2)+2xyx'y'}} \begin{pmatrix} x \\ y \\ -(xx'+yy') \end{pmatrix}. \quad [7]$$

The curvature, $\kappa_T = 1/R_T$, of the fiber given by

$$\kappa_T = \frac{\sqrt{x''^2+y''^2+(y''x'-x''y')^2}}{(x'^2+y'^2+1)^{3/2}}. \quad [8]$$

As outlined above, the vector $e_p$ is assumed to be parallel to the surface normal $e_n$. It follows that:

$$e_p \times e_n \equiv 0 \quad [9]$$

Including the condition of constant fiber curvature (Equation **[8]**) and rearranging give a set of two differential equations in $x, y$, as

$$x'' = \kappa_T \frac{\left(x(1+x'^2)+yx'y'\right)\left(1+x'^2+y'^2\right)}{\left(x^2(1+x'^2)+2xyx'y'+y^2(1+y'^2)\right)^{1/2}}, \quad [10a]$$

$$y'' = \kappa_T \frac{\left(1+x'^2+y'^2\right)\left(xx'y'+y(1+y'^2)\right)}{\left(x^2(1+x'^2)+2xyx'y'+y^2(1+y'^2)\right)^{1/2}}. \quad [10b]$$

The fiber curvature, $\kappa_T$, can be positive or negative (see Supplemental Information for a full derivation). It is helpful to rewrite Equations (**10a**) and (**10b**) in cylindrical coordinates as

$$g'' = \frac{1+g'^2}{g\,(Kg^2-1)}\left(1 - K\,\kappa_T\,g^3\sqrt{1+g'^2}\right), \qquad \text{[11a]}$$

$$\theta' = \frac{1}{g}\sqrt{\frac{1+g'^2}{Kg^2-1}}. \qquad \text{[11b]}$$

The constant $K$ is an integration constant that depends on boundary conditions. Note that both equations above have symmetric solutions where $g(-z) = g(z)$ and $\theta(-z) = -\theta(z)$.

*Boundary conditions*

We consider surfaces of revolutions bounded by two circles with radius $g(\pm L/2) = R$ at the positions $z = \pm L/2$, as sketched in Fig. 2 (upper left corner). The neck radius at $z = 0$ is defined as $g(0) = g_0$ and we search for symmetric solutions with respect to the axis of revolution yielding $g'(0) = 0$. We introduce the fiber angle $\mu_0$ between the $z$-direction and the fiber at the neck ($z = 0$) (see Fig. 1F and Fig. 2), which is given by

$$\cos\mu_0 = 1/\sqrt{1 + g_0^2\,(\theta'(0))^2}, \qquad \text{[12]}$$

With this definition, it follows from Eq. **11b** (taken at $z = 0$) that the integration constant $K$ can be written as

$$K = \frac{1}{g_0^2 \sin^2\mu_0}. \qquad \text{[13]}$$

The boundary conditions sketched in Fig. 1 imply that $g(L/2) = g(-L/2) = R$. In the experimental setting shown in Fig. 1, we set $2L/R = 1.25$, with $g_0/R$ between 0.4 and 0.8. The symmetry condition and the value $g_0$ uniquely define the surface. Each fiber line is then contained in this surface and crosses the neck at $\theta(0)$ with an angle $\mu_0$. Due to axisymmetry we only need to consider fibers for which $\theta(0) = 0$.

Equation **11a** is a second order nonlinear differential equation, and **11b** is a first order non-linear differential equation meaning that for any value of $\kappa_T$ we need a total of two boundary conditions for $g$ and one for $\theta$ as

$$g'(0) = 0,\; g(0) = g_0,\; \theta(0) = 0. \qquad \text{[14]}$$

**A Numerical Study**

It turned out to be more efficient than solving the equations numerically for fixed $\kappa_T$, to only fix $g_0$ and $\mu_0$ over a range of $\kappa_T$. Solutions were accepted that satisfied the additional constraint $g(L/2) = R$, thereby giving the value of $\kappa_T$ in accordance with these conditions. Solving the equations (11) to (13) for surfaces of revolution, we find typically either neck-shaped or barrel-shaped surfaces bounded on the upper and the lower side by disks or radius $R$. The results are summarized in Fig. 2 (see also Figs S1 and S2).

The solutions include expected shapes, such as the sphere, the cylinder and the hyperboloid. The latter is obtained by straight fibers and is found in the graph at $\kappa_T = 0$. Different fibril angles correspond to different neck radii. At the fibril angle of $\mu_0 = 0$, the hyperboloid coincides with a cylinder, as expected. For positive values of $\kappa_T$ solutions tend towards stackings of spherical segments that satisfy equations (11 – 13), except at the joint (see Supplementary Information for more detail). Finally, the upper black curve for $\mu_0 = 0$, corresponds to segments of a torus.

All shapes in Figure 2 will be barrel-like in the upper part where $g_0 > 1$ and neck-like in the lower half, where $g_0 < 1$. All shapes on the left side ($\kappa_T < 0$) correspond to a situation where tensed fibers will create a negative pressure inside the volume (according to Eq.(3)), while for all shapes on the right side of the figure ($\kappa_T > 0$) fiber tension will create a positive pressure. In the case of a growing tissue, negative pressure can be interpreted as a support to the volume increase and, therefore, growing of the tissue, while positive pressure created by the fibers will rather slow down the tissue growth.

For a better comparison with Delaunay surfaces, we redraw Fig. 2 by using the mean curvature at the neck for the x-axis, instead of the fiber curvature (Fig. 3). Note that the mean curvature is not constant for the shapes considered here, while it is constant for the Delaunay surfaces (Fig S4). The Delaunay surfaces are indicated in Fig. 3 by a dotted line. It is quite remarkable that this dotted line is quite close to line corresponding to $\mu_0 = 35°$ in the range of parameters from $g_0$ = 0.4 to $g_0$ = 0.7. This means that surfaces stabilized by fibers with this fibril angle will be close to Delaunay surfaces.

**Implications for the interpretation of experiment data**

We now turn back to the experiments that motivated the current study and compare the data published in Ref. [12] with our model (Fig. 4). Experimental values of neck radius versus curvature were originally interpreted as being compatible with Delaunay surfaces (dashed line Fig. 4A). It turns out however that this data can almost equally well be described by fiber-stabilized volumes of revolution with fiber angles between 30° and 35°. Surprisingly, this is exactly the actin stress-fiber angle that was measured in the experiments (Fig. 4B). Furthermore, the predicted fiber paths within the surface (Fig. 4C) also match remarkably well to the experiments images.

This excellent correspondence between the current model and the experiment seems, therefore, more than coincidental. In addition, there is a fundamental difference in tissue growth constrained by a surface consisting of tensed fibers and growth constrained by a stretched isotropic surface: As evident in Fig. 4C, the mean curvature changes from negative to positive when the neck radius increases beyond approximately $g_0 \approx 0.7$. Based on the Young- Laplace equation (1), this would mean that a surface stress state generates negative pressure, thus enhancing growth, for a neck radius below $g_0 \approx 0.7$. At larger neck radii, this pressure turns positive and would hinder further tissue growth. The experiment data from [13], however does not stop at the neck-radius where pressure would be positive, and no change in growth rate is observed. This implies that the Young-Laplace treatment does not explain the kinetics of growth of the microtissues. One potential explanation is that the tissue surface is highly anisotropic as seen in the twisted actin stress-fibers of Fig. 1C.

According to the treatment in this current paper, the pressure in the volume is proportional to fiber curvature (rather than surface mean curvature), if the surface stress state is generated by fibers, in accordance with Eq. (3). At this point, one needs to refer to Fig. 2: for fibril angles in the neck of $\mu_0$ = 30° or 35°. The fiber curvature $\kappa_T$ remains negative as long as the neck radius does not exceed $g_0 \approx 0.9$. In Fig. 4C, negative values of the fiber curvature correspond to the region below the red line for hyperbolic surfaces with $\kappa_T = 0$. All experimental data points are below the red line in Fig.

4C, which means that in all cases the fiber tension leads to negative pressure inside the tissue, thus enhancing tissue growth.

This observation indicates that cells generating aligned stress-fibers in the near-surface region produce negative pressure inside the tissue more effectively than an isotropic surface stress state. Indeed, for the example studied, an isotropic surface tensile stress state would generate negative pressure only up to a neck radius of $g_0 \approx 0.7$, while aligned stress fibers in the surface provides negative pressures up to a neck radius of $g_0 \approx 0.9$.

At this point, it remains unclear why cells chose an angle of 30° or 35° to develop the stress fibers, but one may speculate, based on the fact that force generation by actin-myosin interaction is always intermittent [28], that in the relaxation phases of the contractile fibers the tissue relaxes to shapes dictated by the surface energy without contractile fibers. Based on the boundary conditions, these shapes would be Delaunay surfaces. It is then not unlikely that cells, when they develop fiber tension, chose an orientation ~30° that mimics best the actual shape but provides additional negative pressure inside the tissue to allow for and even enhance growth. The reason why tissues in [12] choose a particular fiber chirality is also unknown, as our model is equally valid for both possible chirality's. Potentially the inherent handedness of cells [29] may preference one direction over another, and further experiments will be required to investigate this.

**Conclusion**

The well-known Young-Laplace equation can be modified to describe surfaces tensed by parallel fibers instead of isotropic tensile surface stresses. The pressure inside the volume is then proportional to the stress in the fibers and inversely proportional to the radius of curvature of the fibers. An analysis of surfaces of revolution shows that surfaces with constant fiber curvature and in equilibrium with a pressure inside the volume they enclose exist in a wide range of parameters. Comparing the results to previously published experimental data shows that spiral arrangements of stress fibers are more powerful than an isotropic surface tension state in generating negative pressures inside a growing tissue, consequently facilitating tissue growth more efficiently.

**Acknowledgments**

PF acknowledges support by the German Research Foundation within SFB1444 and through the Cluster of Excellence 'Matters of Activity', EXC2025.

**References**

1. Young, T., *An essay on the cohesion of fluids.* Phil. Trans. R. Soc., 1805. **95**: p. 65-87.
2. Laplace, P.S., *Traité de Mécanique Céleste*. Vol. 4. 1805, Paris: Courcier.
3. Thompson, D.W., *On Growth and Form*. 1917, Cambridge: Cambridge University Press.
4. Heisenberg, C.P., *D'Arcy Thompson's 'on Growth and form': From soap bubbles to tissue self-organization.* Mechanisms of Development, 2017. **145**: p. 32-37.
5. Graner, F. and D. Riveline, *'The Forms of Tissues, or Cell-aggregates': D'Arcy Thompson's influence and its limits.* Development, 2017. **144**(23): p. 4226-4237.
6. Dunlop, J.W.C., et al., *The Emergence of Complexity from a Simple Model for Tissue Growth.* Journal of Statistical Physics, 2020. **180**(1-6): p. 459-473.
7. Rumpler, M., et al., *The effect of geometry on three-dimensional tissue growth.* Journal of the Royal Society Interface, 2008. **5**(27): p. 1173-1180.
8. Gamsjäger, E., et al., *Modelling the role of surface stress on the kinetics of tissue growth in confined geometries.* Acta Biomaterialia, 2013. **9**(3): p. 5531-5543.
9. Bidan, C.M., et al., *Geometry as a Factor for Tissue Growth: Towards Shape Optimization of Tissue Engineering Scaffolds.* Advanced Healthcare Materials, 2013. **2**(1): p. 186-194.
10. Callens, S.J.P., et al., *Substrate curvature as a cue to guide spatiotemporal cell and tissue organization.* Biomaterials, 2020. **232**.
11. Kollmannsberger, P., et al., *Tensile forces drive a reversible fibroblast-to-myofibroblast transition during tissue growth in engineered clefts.* Science Advances, 2018. **4**(1).
12. Ehrig, S., et al., *Surface tension determines tissue shape and growth kinetics.* Science Advances, 2019. **5**(9).
13. Delaunay, C., *Sur la surface de revolution dont le courbure moyenne est constante.* J. Math. Pures et appl. Sér 1, 1841. **16**: p. 309-314
14. Wang, L.M. and T.J. McCarthy, *Capillary-bridge-derived particles with negative Gaussian curvature.* Proceedings of the National Academy of Sciences of the United States of America, 2015. **112**(9): p. 2664-2669.
15. Wang, L.W., *Atomistic Approximation of Solid Surface Energy and Its Anisotropy.* Metallurgical and Materials Transactions a-Physical Metallurgy and Materials Science, 2020. **51**(12): p. 6127-6131.
16. Rey, A.D., *Young-Laplace equation for liquid crystal interfaces.* Journal of Chemical Physics, 2000. **113**(23): p. 10820-10822.
17. Cohen, A.E. and L. Mahadevan, *Kinks, rings, and rackets in filamentous structures.* Proceedings of the National Academy of Sciences of the United States of America, 2003. **100**(21): p. 12141-12146.
18. Fischer, F.D., et al., *On the role of surface energy and surface stress in phase-transforming nanoparticles.* Progress in Materials Science, 2008. **53**(3): p. 481-527.
19. Timoshenko, S. and S. Woinowsky-Krieger, *Theory of plates and shells*. 2d ed. Engineering societies monographs. 1959, New York,: McGraw-Hill. 580 p.
20. Jawad, M.H., *Theory and design of plate and shell structures*. 1994, New York: Chapman & Hall. xv, 423 p.
21. Ziegler, F., *Technische Mechanik der festen und flüssigen Körper*. 3 ed. 1998, Wien: Springer Verlag.
22. Goriely, A., *The mathematics and mechanics of biological growth*. Interdisciplinary applied mathematics,. 2017, New York, NY: Springer. xxii, 646 pages.

23. Blaauwendraad, J. and J.H. Hoeffakker, *Structural Shell Analysis: Understanding and Application*. 2014, Dordrecht: Springer Science+Business Media
24. Ferri, J.K., et al., *Elastic nanomembrane metrology at fluid-fluid interfaces using axisymmetric drop shape analysis with anisotropic surface tensions: deviations from Young-Laplace equation.* Soft Matter, 2012. **8**(40): p. 10352-10359.
25. Fischer, F.D., et al., *Tissue growth controlled by geometric boundary conditions: a simple model recapitulating aspects of callus formation and bone healing.* Journal of the Royal Society Interface, 2015. **12**(107).
26. Pressley, A., *Elementary differential geometry*. 2nd ed. Springer undergraduate mathematics series,. 2010, London ; New York: Springer. xi, 473 p.
27. Schlichtkrull, H., *Curves and Surfaces (Lecture Notes for Geometry 1)* 2010, Dptmt. Mathematics, Univ. Copenhagen
28. Vogel, V. and M.P. Sheetz, *Cell fate regulation by coupling mechanical cycles to biochemical signaling pathways.* Curr Opin Cell Biol, 2009. **21**(1): p. 38-46.
29. Wan, L.Q., et al., *Micropatterned mammalian cells exhibit phenotype-specific left-right asymmetry.* Proceedings of the National Academy of Sciences of the United States of America, 2011. **108**(30): p. 12295-12300.

**Figures and Tables**

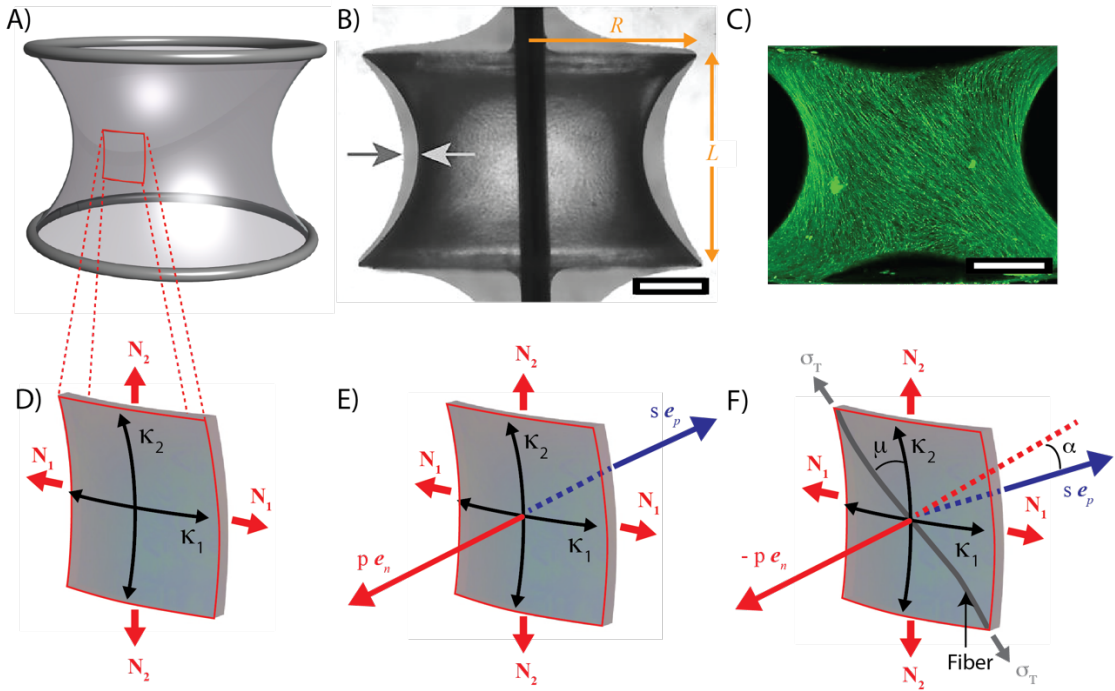

**Figure 1.** Experimental observations and definition of surface coordinates A) Image of a catenoid, i.e. the surface of revolution of a catenary, which satisfies Laplace Young equation for zero pressure difference over the membrane. B) Projection of bone-like tissue grown on a polymeric surface of revolution (Capillary bridge) fixed on a central pin (dark line). The light arrow indicates the boundary of the polymeric surface, the dark arrow indicates the position of the tissue after 30 days growth. $R$ and $L$ are the radius and separation of two circular disks corresponding to the upper and lower boundary of the tissue. C) Projection of a 3D light sheet fluorescence microscopy image of tissue stained for actin (green fibers). Note the strong orientation of the actin stress fibers. D) The Laplace-Young equation (Eq. 1) can be understood by the tension balance over a surface patch, with two principal curvatures, $\kappa_1$ and $\kappa_2$ and membrane forces $N_1$ and $N_2$ (Eq. 2). E) The pressure $p$ of the fluid inside the volume generates a force directed along the normal to the surface $e_n$. For an isotropic membrane, the sum of the membrane forces $s$ is also perpendicular to the surface but pointing in opposite direction $-e_n$. There is equilibrium if $s$ and $p$ have the same magnitude. F) If the local mechanical response of the membrane is only generated by a fiber in the surface with tension $\sigma_T$. Then the resultant local force $s$ lies along $e_p$, within the osculating plane of the fiber, and is not necessarily colinear with $p$ (i.e. $\alpha \neq 0$). Therefore, the conditions for equilibrium of the surface are (1) that the direction of $p$ (that is, $e_n$) lies within the osculating plane of the curve describing the fiber (i.e. the angle $\alpha = 0$) and (2) that the magnitude of $s$ and $p$ are the same, which leads to Eqs. (3 and 4), The fibril angle $\mu$ is measured between the fiber direction and $N_2$. The images in B and C are reproduced from Ref. [12], under the CC BY-NC Licence.

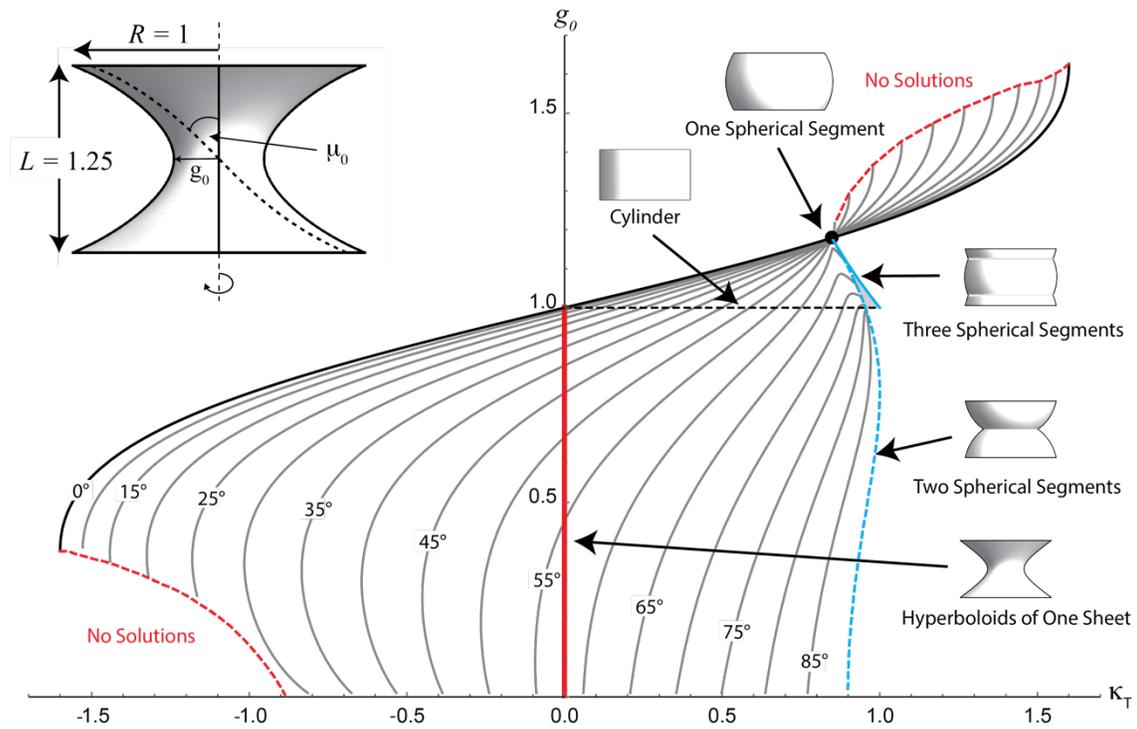

**Figure 2.** Surfaces of revolution satisfying equation(s) **11a, 11b** and the boundary conditions **14** as indicated in the inset top left, characterized by the (constant) curvature $\kappa_T$ of the fibers which stabilize the shape and the neck radius $g_0$. Fibers follow spiraling paths, and the fibril angle at the equator $\mu_0$ is indicated for each set of curvature and neck radius. The inset (top-left) shows one such surface of revolution consisting of fibers of constant curvature that cross the equator with a "microfibril" angle, $\mu_0$. The dashed line on the inset indicates the path of one such fiber. Grey lines in the main diagram give the relationship between $g_0$ and $\kappa_T$ for a fixed fibril angle $\mu_0$ at the equator. The dashed red lines indicate the limits below or above which no solutions can be found. The solid red line shows the range of solutions satisfied by hyperboloids of one sheet, the dashed black line shows the range of cylindrical solutions. The full black circle indicates the solution given by one spherical segment. All solutions with microfibril angles less than 58° pass through this point. The dotted blue line indicates the relationship between $g_0$ and $\kappa_T$ or a stack of two spherical segments, and the solid blue line for a stack of three spherical segments. Note that these solutions are not differentiable at the joint between the spherical segments and can be considered as limit cases. The lower part of the graph ($g_0 < 1$) corresponds to necked structures as shown in the inset. The upper part of the graph describes bulged structures ($g_0 > 1$), akin to the single spherical segment. The light grey region between the dotted blue line and the solid blue line is shown in more detail in the supplementary information (Fig. S2).

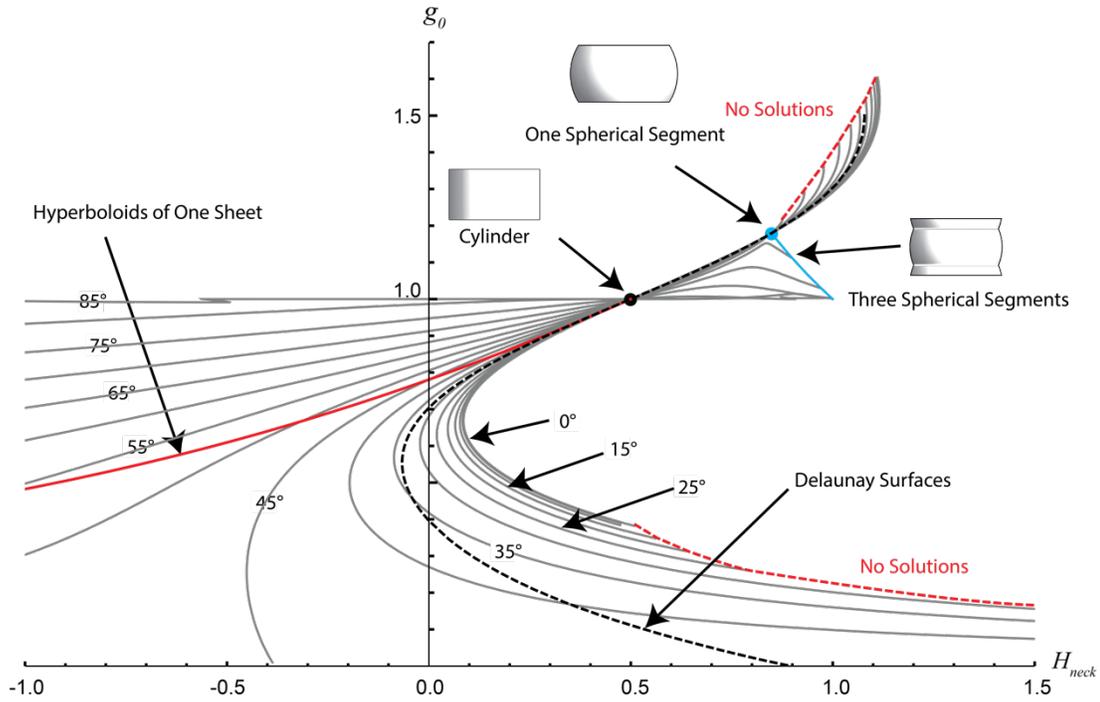

**Figure 3.** Alternative representation of Figure 2, showing the neck radius, $g_0$, versus the mean curvature at the neck, $H_{neck}$, for different "mifrofibril" angle, $\mu_0$, at the equator. Grey lines give solutions in which the microfibril angle at the neck is fixed. The dashed red lines indicate the limits beyond which no solutions can be found. The solid red line shows the range of solutions satisfied by hyperboloids of one sheet, the solid blue line indicates solutions for stacks of three spherical segments. The black dot indicates cylindrical solutions, the blue dot shows the solution for one segmented sphere. The dashed black line shows the relationship between neck radius and mean curvature for Delaunay surfaces that satisfy the boundary conditions. Note that this curve is close to the solutions for $\mu_0$ = 35° for the neck ranging from about $g_0$ = 0.4 to $g_0$ = 0.7.

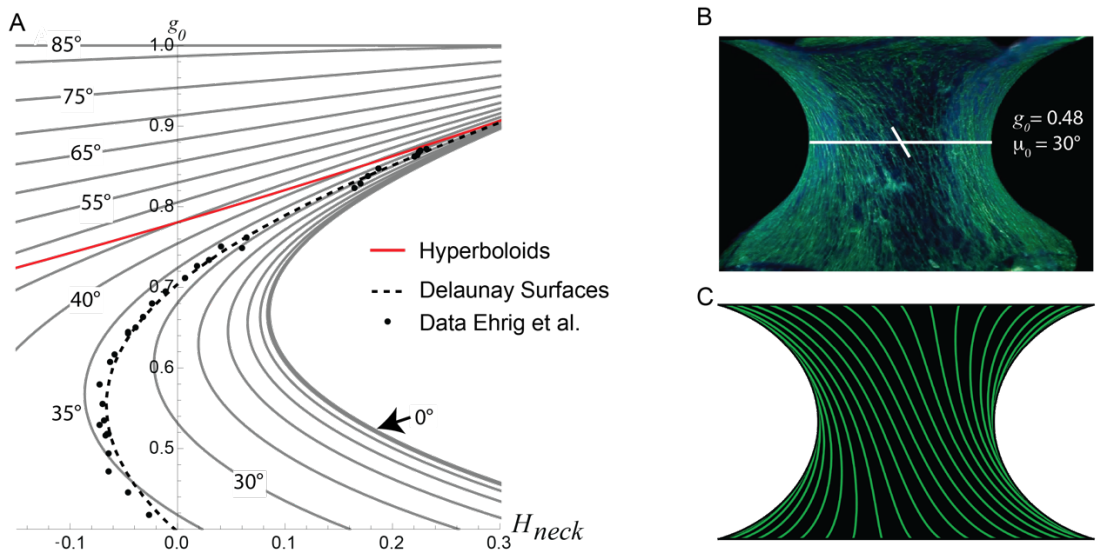

**Figure 4.** Comparison of the model based on anisotropic surface contraction to the experiments of Ref. [12]. Panel A is an enlargement of a portion of Fig. 3 with dots indicating all the tissue growth experiments performed in Ref [12]. It is quite remarkable that these dots are simultaneously close to the line for Delaunay surfaces (dotted line) and to fiber-stabilized volumes with a microfibril angle at the neck between 30° and 35°. Note, no experimental data was produced for neck radii smaller than 0.4, due to the Rayleigh instability which limits how small neck-radii can be in liquid capillary bridges. Panel (B) shows a typical microtissue grown in-vitro under the boundary conditions used in our model calculations. The green coloration is due to actin staining and the microfibril angle in the neck is around 30° to 35° for most of the experiments. Panel C shows a solution to Eq. (11) with predicted fiber paths shown in green and a microfibril angle at the neck of 30°.

**Supplementary Information for**

Anisotropic Young-Laplace-equation provides insight into tissue growth.

Peter Fratzl[1], F. Dieter Fischer[2], Gerald A. Zickler[2], John W.C. Dunlop[3]

[1]Max Planck Institute of Colloids and Interfaces, Department of Biomaterials, Potsdam Science Park, 14476 Potsdam-Golm

[2]Montanuniversität Leoben, Institute of Mechanics, 8700 Leoben, Austria

[3]Morphophysics Group, Department of the Chemistry and Physics of Materials, University of Salzburg, 5020 Salzburg, Austria

*corresponding authors: Peter Fratzl and John W.C. Dunlop

Email: fratzl@mpikg.mpg.de, john.dunlop@plus.ac.at

**This PDF file includes:**

    Supplementary text

    Figures S1 to S4

    Table S1

**Supplementary Information Text**

In the first section of the Supplementary Information, we show the full derivation of equations **10** and **11**. In the second section we explore in more detail the solution space of equations **11a** and **b**, and discuss specific limit cases as of the torus and when the surface of revolution become close to stacks of spherical segments. In the final section we compare solutions of our equations with the well-known Delaunay surfaces of revolution.

**1) Derivations of Equations 10a,b and 11a,b**

We give more details of the derivations in order to help the reader following the text. Equations **6-9** are repeated from the main text as follows:

$$\boldsymbol{e}_p = \frac{1}{\sqrt{x''^2+y''^2+(y''x'-x''y')^2}\sqrt{x'^2+y'^2+1}} \cdot \begin{pmatrix} x''(1+y'^2) - y''x'y' \\ y''(1+x'^2) - x''x'y' \\ -(x'x'' + y'y'') \end{pmatrix}, \quad [6]$$

$$\boldsymbol{e}_n = \frac{1}{\sqrt{x^2(1+x'^2)+y^2(1+y'^2)+2xyx'y'}} \begin{pmatrix} x \\ y \\ -(xx' + yy') \end{pmatrix}, \quad [7]$$

$$\kappa_T = \frac{\sqrt{x''^2+y''^2+(y''x'-x''y')^2}}{(x'^2+y'^2+1)^{3/2}}, \quad [8]$$

$$\boldsymbol{e}_p \times \boldsymbol{e}_n \equiv \boldsymbol{0}. \quad [9]$$

Equation **9** will hold if the following equation is true:

$$\begin{pmatrix} x''(1+y'^2) - y''x'y' \\ y''(1+x'^2) - x''x'y' \\ -(x'x'' + y'y'') \end{pmatrix} \times \begin{pmatrix} x \\ y \\ -(xx' + yy') \end{pmatrix} = \boldsymbol{0}. \quad [\text{S-1}]$$

This implies that:

$$\begin{pmatrix} -(y''(1+x'^2) - x''x'y')(xx' + yy') + (x'x'' + y'y'')y \\ (-(x'x'' + y'y''))x + (x''(1+y'^2) - y''x'y')(xx' + yy') \\ (x''(1+y'^2) - y''x'y')y - (y''(1+x'^2) - x''x'y')x \end{pmatrix} = \boldsymbol{0}. \quad [\text{S-2}]$$

One can show that all terms in this equation are satisfied if the relation below is valid,

$$x''(y + y'(xx' + yy')) = y''(x + x'(xx' + yy')). \quad [\text{S-3}]$$

Rearranging equation **8** as follows:

$$x''^2 + y''^2 + (y''x' - x''y')^2 = \kappa_T^2 (x'^2 + y'^2 + 1)^3. \quad [\text{S-4}]$$

Equation **S-3** can then be solved for $x''$, and $x''$ inserted in equation **S-**4 gives:

$$y''^2 \left( \left( \frac{x+x'(xx'+yy')}{y+y'(xx'+yy')} \right)^2 + 1 + \left( x' - x \left( \frac{x+x'(xx'+yy')}{y+y'(xx'+yy')} \right) y' \right)^2 \right) = \kappa_T^2 (x'^2 + y'^2 + 1)^3. \quad [\text{S-5}]$$

This can be readily solved for $y''$ yielding equations **10a** and **10b**.

$$x'' = \kappa_T \frac{\left(x(1+x'^2)+yx'y'\right)(1+x'^2+y'^2)}{\left(x^2(1+x'^2)+2xyx'y'+y^2(1+y'^2)\right)^{1/2}}, \tag{10a}$$

$$y'' = \kappa_T \frac{(1+x'^2+y'^2)\left(xx'y'+y(1+y'^2)\right)}{\left(x^2(1+x'^2)+2xyx'y'+y^2(1+y'^2)\right)^{1/2}}. \tag{10b}$$

It is useful to solve these equations also in cylindrical coordinates. To do this we rearrange equation **S3** as

$$x\,y'' - y\,x'' = (y\,y' + x\,x')(y'x'' - x'y''). \tag{S-6}$$

Using an intermediate variable, $h = g\theta'$, this equation can be rewritten as

$$(1 + g'^2)\left(\frac{h}{g} + \frac{h'}{g'}\right) - h\left(g'' - \frac{h^2}{g}\right) = 0. \tag{S-7}$$

This can be integrated giving

$$h = g\,\theta' = \sqrt{\frac{1+g'^2}{Kg^2-1}}. \tag{S-8}$$

$K$ is an integration constant that can be linked to the fiber angle at the neck (Equation **13**). Rearranging equation **S-8** directly leads to equation **11b**. Using this result together with Equation **S-4**, rewritten in cylindrical coordinates gives, after some rearranging, equation **11a**.

### 2) Analysis of the solution space of Equations 11a and b

A systematic numerical study of equations **11a** and **11b** was performed to explore the range of geometries of constant fiber curvature surfaces of revolution (Figs. 2, S1 and S2). For negative fiber curvatures the solutions all have profiles with surface curvatures of constant sign, with the profiles "pinching" together at a neck radius of zero. Surfaces with zero fiber curvature are simply ruled surfaces or hyperboloids of one sheet (red-line in Fig. 2). These surfaces have a maximum neck radius of 1 being the cylinder solution (Black dashed line in Fig S2). For cylinders the fiber curvature is given by $\kappa_T = \sin^2\mu_0/R$, with microfibril angles ranging from 0 to 90°. For positive fiber curvatures and small neck radii, the surface curvatures change sign, with local "concave" areas at the neck and smooth "convex" areas above and below (bottom right of Fig S1). For neck radii larger than 1, the solutions become more convex (top right of Fig S1). To further understand the solution space of the equations it is useful to look in more details at the plot of the neck radius and fiber curvature calculated from the numerical solutions of equations **11a** and **11b** (Fig S2). As the fiber curvature increases (towards the right of Fig. S1), we approach limits which correspond to stackings of two or more spherical segments. Although these are not differentiable at the joint between the spherical segments, these limit cases help us understand the geometry of the numerical solutions, which are smoothed versions of the spherical sphere stackings.

A fiber embedded into a spherical surface segment with the same radius of curvature as the fiber curvature, will satisfy equations **11a** and **11b**. This can be seen in the numerical solutions that tend towards spherical solutions for large values of fiber curvature. This means it is possible to construct via stacking of spherical segments new solutions that will also satisfy equations **11a** and **11b**, except at the cusp-like join. This can be better understood in Fig S2B, where the boxed region highlighted in Fig S2A is replotted showing some examples of limit cases consisting of stacked spherical segments. The simplest solution with a spherical segment is the single truncated sphere indicated by the black point in Fig S2B. All numerical solutions, found with fiber angles at the neck less than $\left(\arctan\frac{R}{L/2}\right) \sim 58°$, pass through this point. Numerical solutions with higher microfibril angles at the neck show a more complicated relationship between neck radius and fiber curvature (Fig S2B). The neck radius decreases with increasing fiber curvature (see e.g. the set of solutions for μ = 65°) and even starts oscillating for solutions close to the cylinder (black dashed lines). The relationship between neck radius and the fiber (or sphere) radius of curvature of these surfaces can be solved exactly for a given L and R (Table S1). For double spheres the curvature is a function of neck radius, $g_0$, (Fig S3 B-D). For surfaces constructed with 3 or more sphere segments (Fig S3 E-G) the surfaces can be parameterized by $l_s$, which is the height of the largest sphere segment.

The black line in Fig S2, corresponding to $\mu_0 = 0°$, is the limit case which can be described by a segment of a torus. Combining the standard equation of a torus with the boundary conditions at $z = L/2$ gives a relation between neck radius and curvature as

$$g_0 = R \pm \sqrt{\frac{1}{\kappa_T^2} - \frac{L^2}{4}} \pm \frac{1}{|\kappa_T|}. \quad\quad\quad \textbf{[S-9]}$$

The limit values of curvature corresponding to the end points of the black line (Fig S2) are given by the dashed lines $g_0^{lim} = 1 \pm 2R/L$ and $\kappa_T^{lim} = \pm 2R/L$.

### 3) Comparison with Delaunay Surfaces

The profiles calculated using equations **11a** and **11b**, are compared to Delaunay surfaces calculated using the equations of Gillette and Dyson [1] and plotted in Fig. S4 for different neck radii. The profiles are surprisingly similar. Differences become only apparent when the mean curvature of the surfaces are compared. This indicates that fiber supported surfaces can closely approximate Delaunay constant mean curvature surfaces, however with different signs of internal pressure as discussed in the main text.

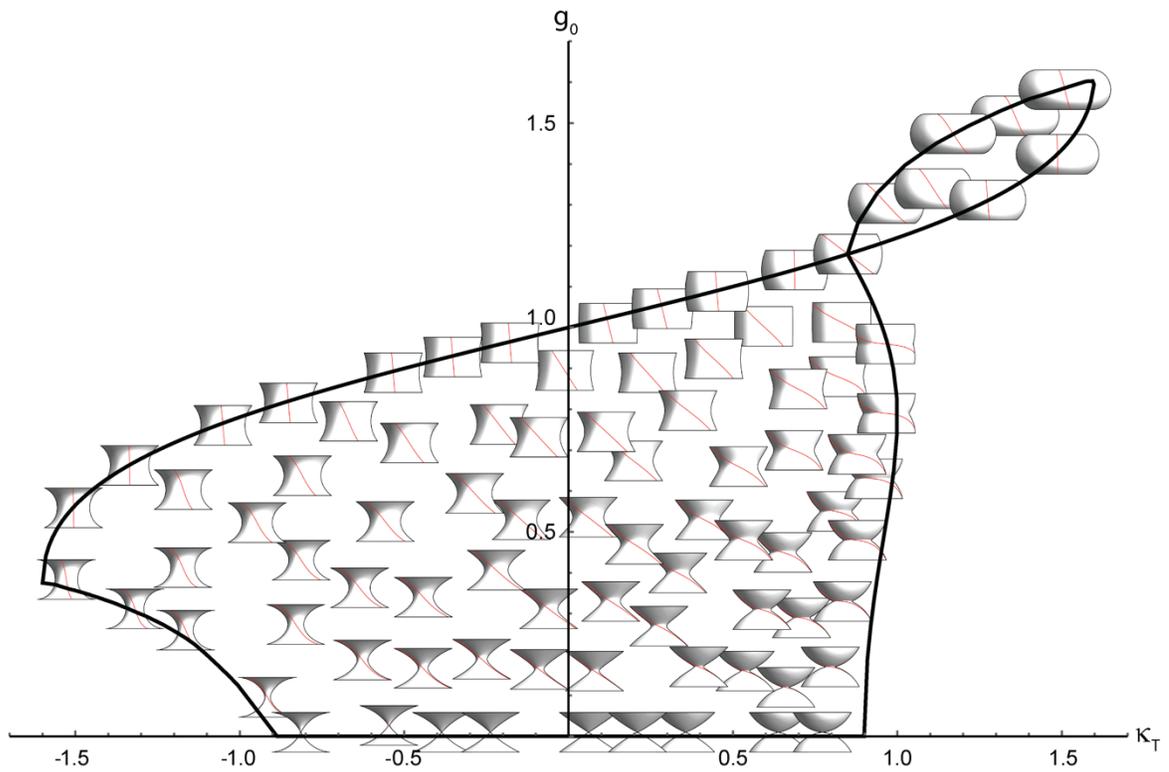

**Figure S1** – Images of a selection of surfaces of revolution and their constant curvature fiber generators (red), satisfying equations **11a** and **11b** are plotted as a function of their neck radii and fiber curvature (L = 1.25, R = 1). The outer boundary (**blacklines**) describe the boundaries of the solution space in Fig **2** in the Main Manuscript.

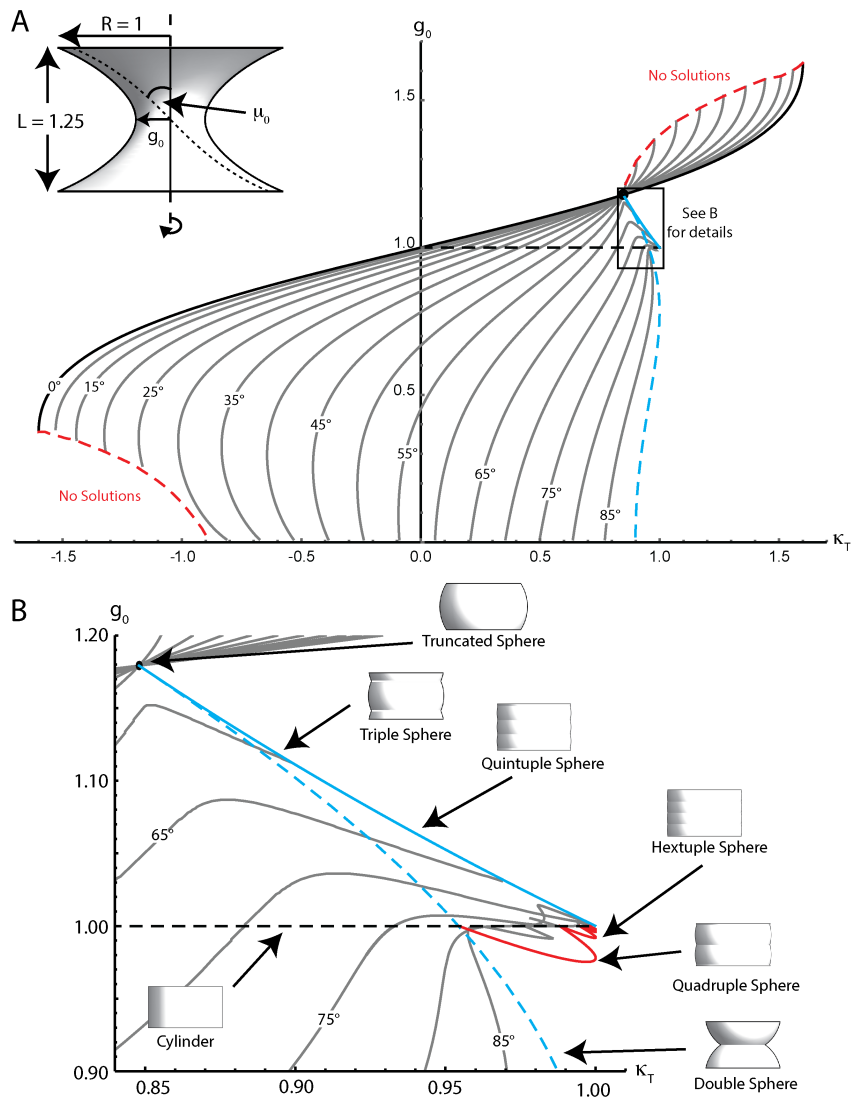

**Figure S2** – Numerical solutions of neck-radius g0, versus fiber curvature $\kappa_T$ A) The normalized neck radius versus the fiber curvature for surfaces of revolution satisfying equation(s) **11a** and **11b** (L =1.25, R = 1). Grey lines give the numerical solutions in which the microfibril angle at the neck is fixed at the angle indicated. The dashed red lines indicate the limits below or above which no solutions can be found. The dashed black line shows the range of cylindrical solutions. The black circle is the truncated sphere solution that satisfies equations **11a** and **11b**. The dotted blue line indicates the double truncated sphere solutions. The solid blue line indicates the parameters of triple truncated sphere solutions. The inset shows an exemplary solution B) A zoom of the boxed region from part A), shows in addition solutions for quadruple and hextuple spheres (Red) and a detail of the numerical solutions for microfibrils of 60° to 85°. Note surfaces made of an odd number of sphere segments (triple and quintuple sphere surfaces) have parameters lying on the same solid blue line.

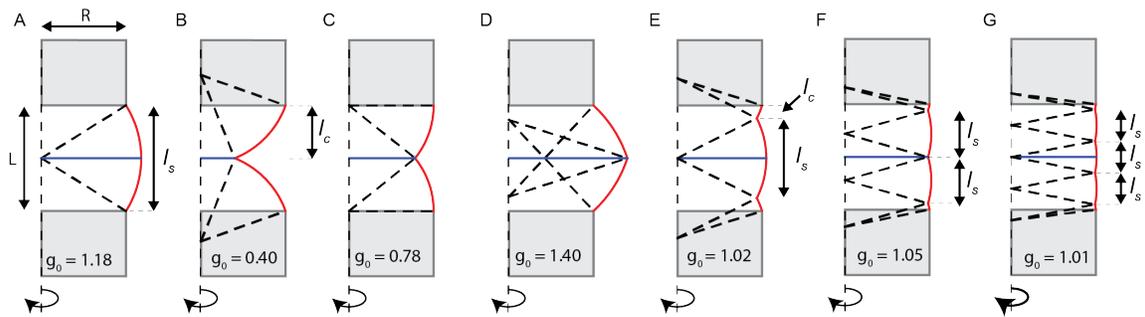

**Fig. S3** – Profiles of stacked spherical segment solutions according to equations **11a** and **b**, for different neck radii. A) Truncated sphere, B), C), and D) Double sphere for different neck radii, E) Triple sphere, F) Quadruple sphere, and G) Quintuple sphere. Red lines indicate the profile curve, blue lines give the neck radius ($g_0$), and the dashed lines indicate the direction to the center of curvature of the different segments of the profiles. For surfaces constructed with 3 or more segments (E-G), the height of the largest sphere segment, $l_s$, is highlighted, $l_c$ is the height of the smallest sphere segment.

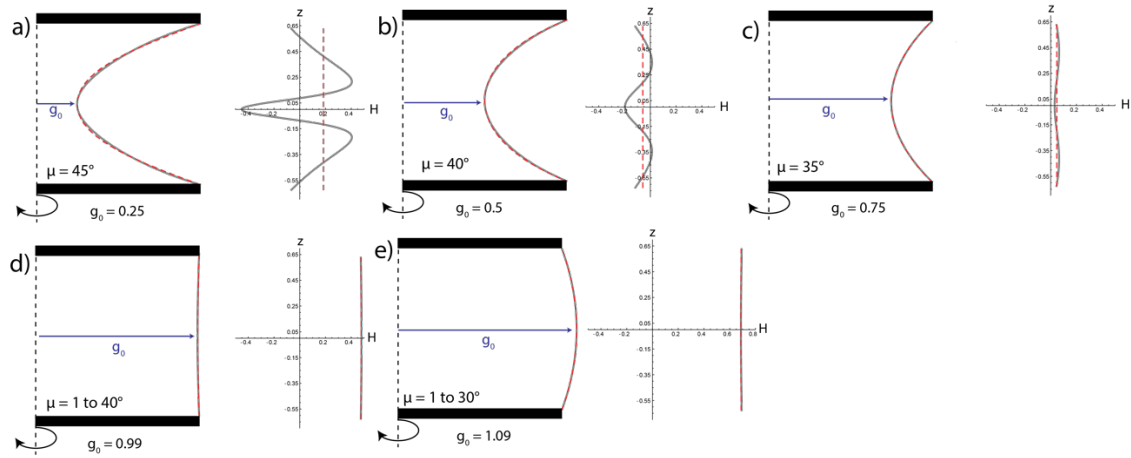

**Fig. S4** – a-e (left) Profiles for different neck radii of numerical solutions according to Equations **11a** and **11b** in gray, compared to Delaunay surfaces in red. For each profile the mean curvature as a function of z is also plotted for both types of surface.

| $n$ | $l_s$ | $l_c$ | $g_0$ | $(R_T = 1/\kappa_T)$ |
|---|---|---|---|---|
| 0 | $L$ | 0 | $\sqrt{R^2 + \frac{1}{4}L^2}$ | $g_0$ |
| 1 | 0 | $\frac{L}{2}$ | $0 \leq g_0$ | $\sqrt{R^2 + \frac{1}{L^2}(R^2 - g_0^2 + L^2)^2}$ |
| 2 | $\frac{L}{2} \leq l_s \leq L$ | $(L - l_s)/2$ | $\sqrt{R^2 + \frac{1}{4}(2l_s - L)^2}$ | $g_0$ |
| 3 | $\frac{L}{3} \leq l_s \leq \frac{L}{2}$ | $(L - l_s)/2$ | $\sqrt{R^2 + \frac{1}{4}(3l_s - L)^2}$ | $\sqrt{R^2 + \frac{1}{4}(3l_s - L)^2 - \left(\frac{l_s}{2}\right)^2}$ |
| Even | $\frac{L}{n+1} \leq l_s \leq \frac{L}{n-1}$ | $(L - l_s)/2$ | $\sqrt{R^2 + \frac{1}{4}(nl_s - L)^2}$ | $g_0$ |
| Odd | $\frac{L}{n+1} \leq l_s \leq \frac{L}{n-1}$ | $(L - l_s)/2$ | $\sqrt{R^2 + \frac{1}{4}(nl_s - L)^2}$ | $\sqrt{R^2 + \frac{1}{4}(nl_s - L)^2 - \left(\frac{l_s}{2}\right)^2}$ |

**Table S1** – Relationship between the number $n$ of stacked spheres, neck radius $g_0$ and fiber curvature $R_T = 1/\kappa_T$ for different stack sphere heights.

**SI References**

1. R. D. Gillette, D. C. Dyson, Stability of fluid interfaces of revolution between equal solid circular plates. *Chem. Eng. J.* 2, 44-54 (1971).